# NEW OUTCOMES IN MUTATION RATES ANALYSIS

Valentina Agoni
valentina.agoni@unipv.it

**The GC-content is very variable in different genome regions and species but although many hypothesis we still do not know the reason why. Here we show that a relationship exists with the mutation rate, in particular we noticed a new recurrence in the amino acids coding table. Moreover we analyze recombination frequency taking into account Single Nucleotide Polymorphisms hourglass distribution.**

## 1. Error correction and the GC-content

The reason of the high variability in the GC-content between species remains unclear. As mentioned by [1], many hypothesis have been postulated during the years by many groups, such as UV resistance [2][3], thermal adaptation [4][5], directional mutation pressure [6], metabolism[7], the length of coding sequences [8][9], nitrogen fixation [10], aerobiosis [11], environment pressure [12], genome size [13] and DNA polymerase III [14].

What we know for sure is that chromosomal bands reveal different compositions in G+C and that the gene density in the GC-richest fraction of the human genome is known to be higher than in the poorer GC regions [15].

Amino acids coding table evolved to minimize mutations therefore this codons organization suggests C↔T and A↔G mutations to be more frequent.

In his storic paper "Codon—anticodon pairing: The wobble hypothesis" [16], Crick suggests that, due to the degeneracy of the genetic code, wobble base pairs are fundamental for the proper translation of the genetic code. So that while the first two positions of the triplet are critical, there may be some wobble in the pairing of the third base. In other words, the third base is considered to be less discriminatory than the other two bases.

However, if we look at the amino acids coding table (Figure 1) and we focus on amino acids encoded by two different triplets (Phe, Tyr, His, Gln, Asn, Lys, Asp, Glu, Cys, Arg), we can notice that the triplets can be divided into 2 categories: codons which third nucleotide can be U or C (pink circles), and codons which third nucleotide is A or G (purple circles).

This cannot be due to casuality. Here we hypothesis error correction to explain the dilemma.

The determinant factors could be the number of bonds (each couple includes a nucleotide that forms 3 bonds and a nucleotide that forms 2) or the nucleotide formula (purines or pirimidines).

Moreover, we hypothesize polymerase mutations of a 3-bonds-forming-nucleotide to a 2-bonds-forming-nucleotide – from now we will call this type of 3→2 – to be not only the more conserved but the more probable too.

The neutral theory of molecular evolution, introduced by Kimura between '60s and '70s [17], states that the vast majority of conserved mutations are neutral.

Sueoka quantitative theory of directional mutation pressure [18] indicates that the major cause for a change in DNA G + C content of an organism is the large excess of GC→AT mutations over AT→GC mutations.

**Figure 1.** The amino acids coding table.

We hypothesize 3➔2 mutations to be more probable also because coding regions (which are less tolerant for casual not synonymous mutations) have an higher GC-content. We can hypothesis this to be not limited to the 3$^{rd}$ nucleotide.
It will be probably good to consider this phenomenon for new drugs resistance. Bacteria subjected to a high evolutionary pressure, for example related to antibiotics resistance, should have a lower GC-content. Imagine to study a new drug active against bacteria, knowing if the mutation responsible for resistance belongs to the more probable class it can makes a difference. Another important application can be the cancer predisposition probability.

## 2. Role of chromosomal recombination in new species generation

The frequency of Single Nucleotide Polymorphisms (SNPs) along chromosomes follows the typical 'hourglass distribution' [19] (Figure 2). This typical distribution suggests — being centromeres a physical constrain to crossing over — the more variable genes in the different species to be at the end of chromosome arms, then the SNPs number to be proportional to the length of the arm. As a consequence this can be the main difference between species: the variability of genes more than the protein characteristics.
Moreover we know that euchromatic regions undergo crossing over with an high probability [20].
It is known that CENP-A, a centromere protein, is able to identify centromeres by itself (with or without epigenetic modifications of centromeric DNA).
We hypothesize that the process of creating a new specie starts from a non-homologous recombination that leads to centromere repositioning.
In this case the discriminant thing between species could be not only the sequence but also the number of SNPs in a particular gene.

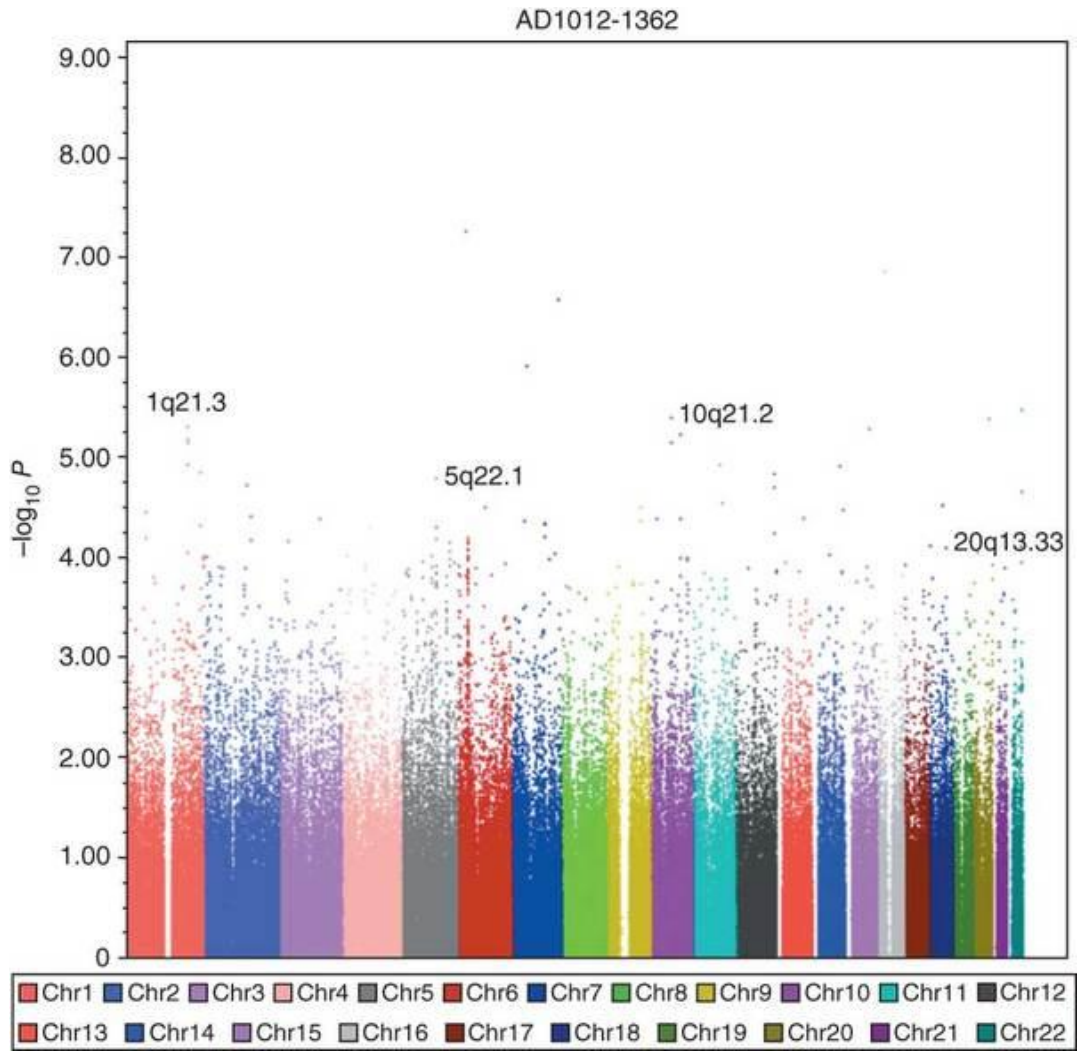

**Figure 2.** SNPs typical hourglass distribution along chromosomes [19].

The minimum length of the annealing region required for crossing over is 25bp while the optimum is about 400bp. In fact, the recombination rates increase three to four orders of magnitude as homology rises from 25 to 411bp [21].
Here we use the combinatorial calculus dispositions with repetitions formula:

$$D_{n,k} = n^k$$

where D represents the possible sequences of k objects extracted from a set of n objects, to calculate the number of different possible combinations of the 4 nucleotides 400bp long. In our case k=4= TCGA and n=400 therefore we obtain $400^4 = 2.5 \ast 10^{10}$ combinations. This number is smaller respect to human genome dimension ($3 \ast 10^9$ bp). This means that every combination appears once (on average) in the human genome or, in other words there are $10^7$ different sequences of 400bp.
We know that in principle even a single mutation in the minimum annealing region of 25bp is sufficient to block recombination. On the other hand there is a high probability to find another region to match with ($10^7$).
Although SNPs distribution is not linear along the chromosome (consider euchromatic regions and 'hourglass distribution') we can make an appoximation taking into account a mutation rate of $10^{-6}$ (as the polymerase error rate is) [22].

This permits to calculate the probability that a mutation occurs in a 25bp sequence:

$$25/10^6 = 2.5*10^{-5}.$$

The reciprocal number represents the dimension (number of items along a certain time interval) of an isolated population before it becomes a new species. We can say species are the discrete elements of the continuous evolutive process.

In conclusion we can say that both SNPs and chromosomal recombination play a role in evolution. Evolution is the result of two opposite forces: error correction and mutations. SNPs and chromosomal recombination act in two very different ways.